\documentclass[12pt]{article}
\usepackage[font=small]{caption}
\usepackage{graphicx}
\usepackage{multirow}
\makeatletter
\makeatletter

\usepackage{xcolor}
\usepackage{newtxtext,newtxmath}

\usepackage{graphicx}

\usepackage[letterpaper,margin=1in]{geometry}

\usepackage{lineno} 

\renewenvironment{abstract}
	{\quotation}
	{\endquotation}

\date{}

\makeatletter
\renewcommand{\fnum@figure}{\textbf{Figure \thefigure}}
\renewcommand{\fnum@table}{\textbf{Table \thetable}}
\makeatother

\usepackage{scicite}

\usepackage{url}

\def\scititle{Microcomb-driven large-scale fully connected quantum network
}
 \title{\bfseries \boldmath \scititle}

\author{names}
\author{
	Fang-Xiang~Wang$^{1,2,3}\dagger$,
	Sheng-Teng~Zheng$^{1,2,3,4}\dagger$,
    Long~Huang$^{5,6}\dagger$,
	\and
 	Guo-Wei~Zhang$^{1,2,3}\dagger$,
	Guang-Shu~Wang$^{1,2,3}$,
	Wen-Jing~Ding$^{1,2,3}$,
    \and
    Ze-Hao~Wang$^{1,2,3}$,
    Shuang~Wang$^{1,2,3,4}$,
	Zhen-Qiang~Yin$^{1,2,3,4}$,
    Chang-Ling Zou$^{1,2,3,4}$,
    \and
    Brent E. Little$^{5}$,
    Guochao~Wang$^{8}$,
    Lingxiao Zhu$^{8}$,
    Guang-Can~Guo$^{1,2,3,4}$,
    \and
    Weiqiang~Wang$^{7}\ast$,
    Wenfu~Zhang$^{5,6}\ast$,
    Wei~Chen$^{1,2,3,4}\ast$,
    Zheng-Fu~Han$^{1,2,3,4}$
	\and\small$^{1}$Laboratory of Quantum Information, University of Science and Technology of China, Hefei 230026, China
	\and
	\small $^{2}$Anhui Province Key Laboratory of Quantum Network, \and\small University of Science and Technology of China, Hefei 230026, China
	\and
	\small$^{3}$CAS Center for Excellence in Quantum Information and Quantum Physics,\and\small University of Science and Technology of China, Hefei 230026, China
	\and
	\small$^{4}$Hefei National Laboratory, University of Science and Technology of China, Hefei 230088, China
	\and
	\small$^{5}$State Key Laboratory of Ultrafast Optical Science and Technology, Xi'an Institute of Optics and Precision Mechanics,
    \and \small Chinese Academy of Sciences, Xi’an 710119, China
	\and
	\small$^{6}$University of Chinese Academy of Sciences, Beijing 100049, China
	\and
	\small$^{7}$ School of Electronic Information and Artificial Intelligence,\and \small Shaanxi University of Science and Technology, Xi'an 710021, China
	\and
	\small$^{8}$ College of Intelligence Science and Technology, National University of Defense Technology, 
    \and\small Changsha 410073, China
	\and\small$^\ast$W. W.: wangwqopt@163.com; W. Z.: wfuzhang@opt.ac.cn; W. C.: weich@ustc.edu.cn.
	\and
	\small $^\dagger$ These authors contributed equally to this work.
}

\begin{document} 

\maketitle
\captionsetup[figure]{name={Fig.},labelsep=period,singlelinecheck=off,justification=raggedright} 
\begin{abstract} \bfseries \boldmath
Fully connected quantum networks enable simultaneously connecting every user to every other user and are the most versatile and robust networking architecture. However, the scalability of such networks remains great challenge for practical applications. Here we construct a large-scale fully connected quantum network founded on two-photon Hong-Ou-Mandel (HOM) interference, where user-to-user security is guaranteed even with untrusted network provider. Using integrated soliton microcomb (SMC) and photonic encoding chips, we realize precise massive parallel frequency generation and locking, high-visibility HOM interferences and measurement-device-independent (MDI) quantum key distribution. The proposed architecture enables a 200-user fully connected quantum network over 200 kilometers with strict information-theoretic security via untrusted network provider. The implemented networking architecture paves the way for realizing large-scale fully connected MDI quantum networks across metropolitan and intercity regions.

\end{abstract}

\section{Introduction}
\label{section:introduction}
Quantum key distribution (QKD) \cite{bennett2014_bb84} monitors eavesdropping by utilizing fundamental principles of quantum mechanics and enables information-theoretically secure communication even under quantum computing attacks \cite{gllp2004,wangxb2005_decoy,lo2005_decoy,xufh2020_rmp}. QKD has made significant achievements over recent years and demonstrated secure key distribution abilities over long distance \cite{lucamarini2018_tf,minder2019_tf,wangs2022_tf,liuy2023_1000km} and field network from fiber to space-to-ground quantum network \cite{frohlich2013_bb84_network,wangs2014_bb84_network,chenya2021_space-to-ground_network}. The security of such networks depends on the trustworthiness of network providers and weakens the quantum-guaranteed security. By leveraging two-photon Hong-Ou-Mandel (HOM) interference \cite{hong1987_hom} and post-selected Bell state measurement (BSM), measurement-device-independent QKD (MDI-QKD) closes all detection loopholes of the QKD system, thus, offers a higher security level in practical implementation \cite{lo2012_mdi,braunstein2012_mdi}. However, although MDI-QKD has achieved long-distance secure key distribution \cite{yinhl2016_404km_mdi}, to implement a large-scale MDI-QKD network is still a formidable task, as an N-user point-to-point MDI-QKD network requires precise frequency locking between N remote laser sources and only the three-node proof-of-principle network were verified \cite{tangyl2016_mdi_network,roberts2017_mdi_network,fan-yuan2021_mdi_network,wangc2022_mdi_network}.

The most versatile and robust networking architecture should realize simultaneously interconnecting among every terminal user, namely, the fully connected quantum network \cite{wengerowsky2018_entanglement_fully_network}. However, the fully connected network is resources consuming and suffers from ultra-low secure key rate \cite{wengerowsky2018_entanglement_fully_network,joshi2020_entanglement_fully_network,liux2022_entanglement_fully_network,huangy2025_entanglement_fully_network}. By executing the MDI-QKD protocol with two-photon HOM interference, the end-to-end secure key rate of a fully connected quantum network could be improved dramatically without trusting the network providers. By placing the costly receiver with single-photon detectors at network provider, MDI-QKD enables a high-security network with low user costs. However, an N-user fully connected MDI quantum network requires precise frequency locking between $\mathcal{O}(N^2)$ laser pairs with parallel frequency channels. As high precise commercial laser sources are typically locked into particular transition lines of atomic or molecular gases and are only usable for some particular wavelengths, it gives rise to unfeasible technical challenges to establish an MDI fully connected quantum network regarding network scale and costs.

Recently, integrated photonic technique has shown advantages in reducing quantum system size, weight and power consumption, as well as production cost \cite{sibson2017_chip_bb84qkd,wangjw2018_chip_quantum_computing,weikj2020_chip_mdiqkd,paraiso2021_chip_bb84qkd,liw2023_chip_bb84qkd}. On the other hand, integrated photonic circuits enable compact frequency multiplexing technique to increase communication bandwidth and expand the optical network scale in both classical \cite{marin-palomo2017_optical_communication} and quantum systems \cite{wangs2014_bb84_network}. Especially, on-chip dissipative Kerr \cite{herr2014_soliton,stern2018_onchip_soliton,kippenberg2018_soliton_review} solitons show remarkable potential in realizing massively parallel high-speed classical \cite{marin-palomo2017_optical_communication,shuhw2022_soliton_optical_communication} and quantum backbone links \cite{wangfx2020_soliton_qkd}. A single-soliton microcomb (SMC) chip can offer hundreds of parallel frequency sources with frequency and phase locking. By locking the seed (pump) laser and repetition rate, massively parallel HOM interferences have been proved between independent SMC chips, which indicated the feasibility to realize a fully connected MDI quantum network using SMCs \cite{huang2025_soliton_hom}.

Here, we demonstrate a high-performance fully-connected MDI-QKD network using integrated photonic technique. The network allows 200 frequency channels connecting and sharing information with each other simultaneously. Each user only requires a single seed laser with precise frequency locking. All seed lasers are with the same wavelength and hence perfectly solves the massive parallel frequency locking challenge. By utilizing silicon-photonic-chip-based MDI-QKD transmitters operating at 2.5 GHz, the network considers the finite-size effect and achieves an average secure key rate of 62 bps at 200 km for each user-to-user connection. The user scale and secure key rate of our network are improved by one and three orders of magnitude than those of previous reported fully connected quantum networks, respectively. The implemented networking architecture is a groundbreaking for practical large-scale fully connected quantum network with high security level.

\section{Results}
\label{results}

\begin{figure}[htbp!]
	\centering
	\includegraphics[width=\textwidth]{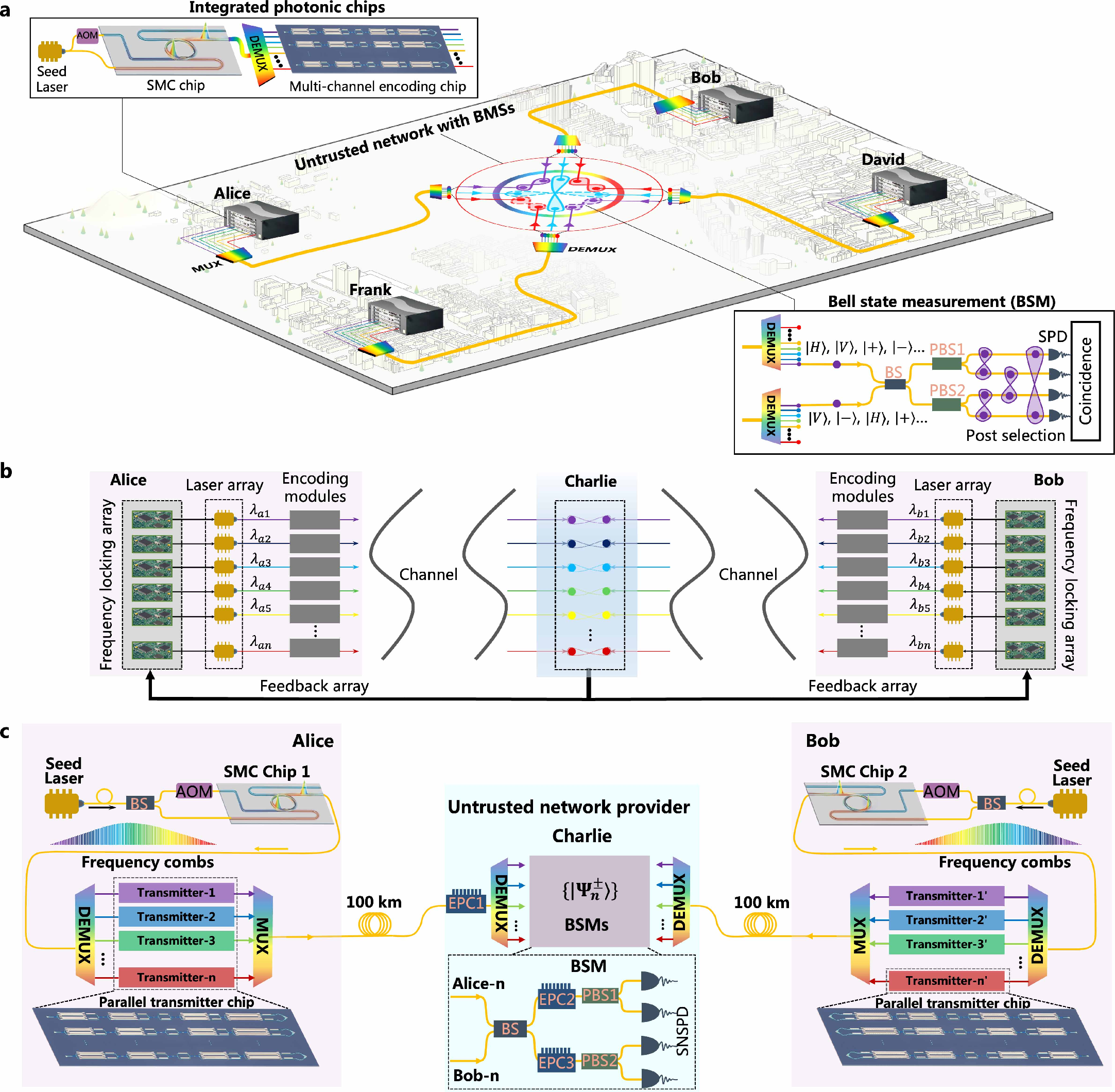} 	
	\caption{\textbf {Fully connected quantum network with untrusted network provider.}
		\textbf{a}, Schematic fully connected MDI-QKD network (four users as an example), in which all terminal users could connected to each other simultaneously without trusting the network provider and each user requires only a frequency-locked seed laser source. \textbf{b}, The typical frequency-locking laser arrays for massively parallel MDI-QKD network, which is a formidable technical challenge. \textbf{c}, Experimental setup of the sub-system of the massively parallel fully connected MDI-QKD network using SMC and QKD transmitter chips over 200 km.}
	\label{fig1:network_concept} 
\end{figure}

A massively parallel fully connected MDI-QKD network is constructed using SMC and MDI-QKD transmitter chips, as shown in Fig. \ref{fig1:network_concept}a. Frequency-locked SMCs are employed as multi-channel carriers and quantum states are prepared with multi-channel encoding transmitter chips at every QKD network terminals. Then hundreds-of-channels quantum states are sent to the untrusted quantum network. The network distributes these channels to different connection links and executes the post-selected Bell state measurements (BSMs). All users realize secure information sharing simultaneously after the network provider publicly announces the BSM results. According to the fundamental principles of quantum mechanics, the quantum relay in the network cannot obtain any information shared between end users during Bell state measurements \cite{lo2012_mdi,braunstein2012_mdi}. We utilize SMC and transmitter chips to implement the massively parallel MDI-QKD, where the seed laser frequency and repeating-rate are locked to align all the soliton comb lines. The frequency locking of SMC chips are fulﬁlled locally and demands no remote operation from network provider or other end users. The independent frequency-locked SMC chips can provide hundreds of available frequency comb teeth pairs with high HOM interference visibilities for fully connected MDI-QKD networks. Hence, the proposed architecture here significantly simplifies the implementation complexity and expands the user scale of a fully connected quantum network, while the typical frequency locking method for $\mathcal{O}(N^2)$ laser pairs with different wavelengths faces enormous technical challenges (as shown in Fig. \ref{fig1:network_concept}b) and only three-node network were realized \cite{tangyl2016_mdi_network,pittaluga2025_tf_network}.

Figure \ref{fig1:network_concept}c shows the sub-system of the massively parallel fully connected MDI-QKD network based on independent SMC and transmitter chips over 200 km. At the end user side (Alice and Bob), part of the seed laser acts as the pump laser and coupled into the micro-ring resonator (MRR) on the SMC chip to generate wide range frequency comb source. The remaining part of the seed laser is frequency-shifted by an acousto-optical modulator (AOM) and acts as auxiliary laser to balance the thermal inner the MRR. The spectrum range of a single SMC chip is over 100 nm, which can provide more than 200 frequency channels with high signal-to-noise ratio (SNR, $>$20dB) and large frequency interval ($\sim$49 GHz). The frequency comb teeth are then demultiplexed by an arrayed-waveguide grating (AWG) filter and distributed to the transmitter chip. The transmitter chip chops the frequency teeth into narrow pulses, then attenuates and encodes the pulses into single-photon-level polarization states. The two bases of polarization states are, ${Z:=}\{|H\rangle, |V\rangle\}$ and ${X:=}\{{|+\rangle=\frac{1}{\sqrt{2}}(|H\rangle+|V\rangle)}, {|-\rangle=\frac{1}{\sqrt{2}}(|H\rangle-|V\rangle)}\}$, separately, where $|H\rangle$ ($|V\rangle$) is the horizontal (vertical) polarization. The encoded pulsed photons are then transmitted to the untrusted quantum network provider Charlie over a 100-km fiber channel, corresponding to 200 km of the user-to-user distance. The polarization drifts of fiber channels are compensated using electrically driven polarization controllers (EPCs). Charlie demultiplexes received photon pulses and performs post-selected BSM detection with two-photon coincidence. The BSM only post selects $|{\Psi}^{\pm}\rangle=\frac{1}{\sqrt{2}}|HV\rangle\pm|VH\rangle$, which consists of a non-polarizing beam splitter (BS, for HOM interference) and two polarizing BS (PBS, for polarization measurement). Charlie announces all two-photon coincident measurement results publicly. Alice and Bob then successfully share a string of secure keys by executing post-processing procedure according to Charlie's announcements. The security of key sharing process is independent on Charlie's loyalty. Hence, by flexibly allocating frequency channels, the proposed architecture can be used to realize both massively parallel high-speed backbone links and fully connected quantum network with untrusted quantum network providers.

\subsection{Local locking of independent SMC chips}
\label{subsection:soliton}
\begin{figure}[htbp!] 
	\centering
	\includegraphics[width=\textwidth]{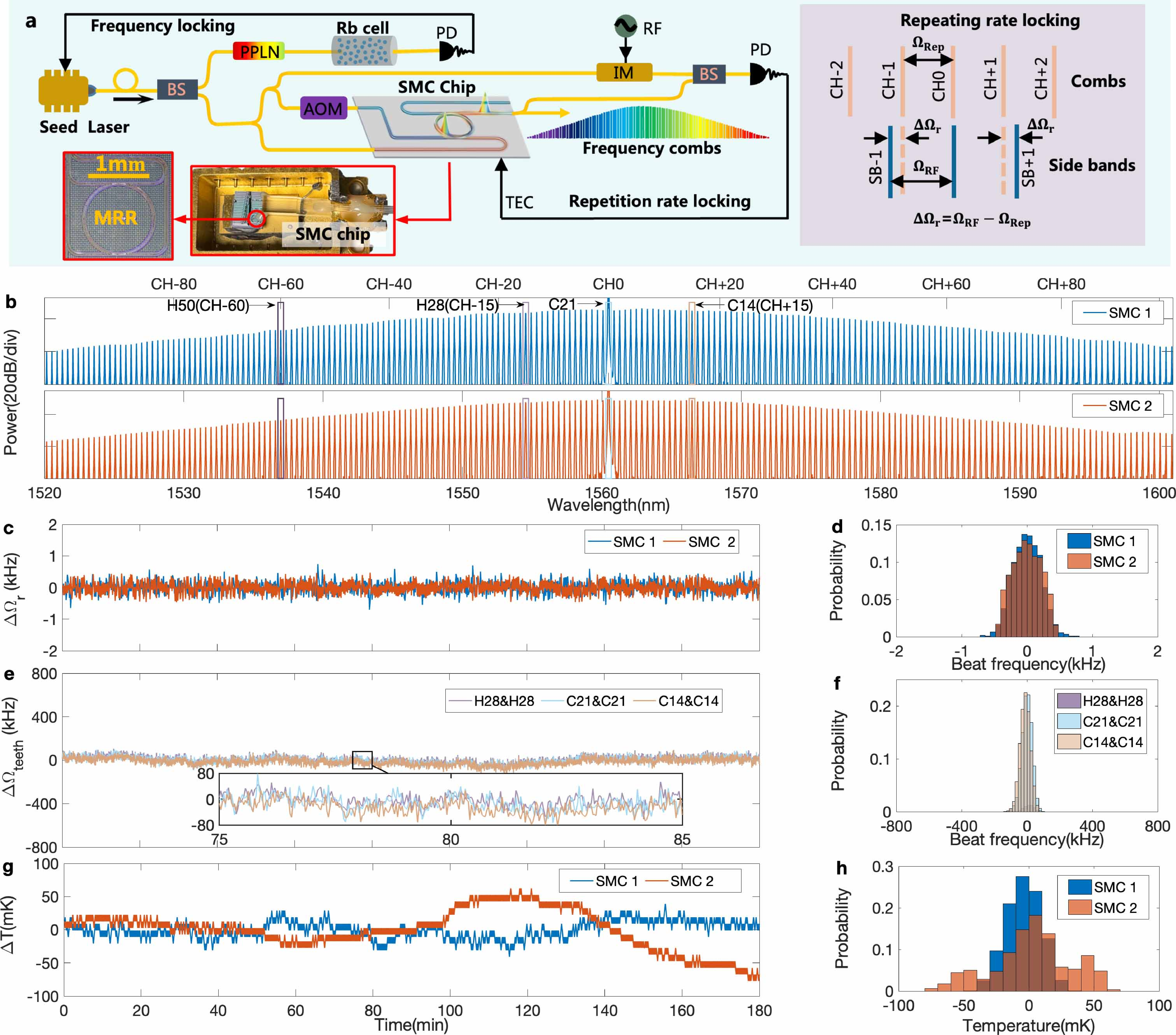} 	
	\caption{\textbf{Locally fully stabilized SMC chips.}
		\textbf{a}, Schematic diagram of SMC local frequency locking setup. The seed laser is locked to a local Rubidium cell. The repetition rate of the SMC chip is locked to a 49-GHz local radio frequency (RF) reference via an intensity modulator (IM). \textbf{b}, Optical frequency spectra of two independently locked SMC chips. \textbf{c-d}, The concurrent repetition rate fluctuations in realtime and statistics of two independent SMCs within 3-hour running. \textbf{e}, The real-time beat frequencies of three pairs of comb teeth within the concurrent time span and \textbf{f} their corresponding statistical distributions. \textbf{g-h}, The monitored real-time operation temperature fluctuation and corresponding statistical distributions of the independent repetition rate locked SMC chips.} 
	\label{fig2:soliton} 
\end{figure}

The photon frequency alignment at diﬀerent transmitters is the fundamental requirement of MDI-QKD  which means that all the independent SMC chips in the network should be locked to the same reference. Figure \ref{fig2:soliton}a presents the local frequency locking scheme of the SMC chip in the proposed fully connected MDI-QKD network. The two degrees of freedom of on-chip soliton microcomb are the seed laser frequency and repetition rate, which, respectively, determine the center frequency (CH0) and repetition rate $\rm\Omega_{Rep}$ of the frequency comb. The parallel seed lasers are locally locked to the D2 transition line of Rubidium cells (${}^{87}$\textbf{Rb}, $\rm5^2S_{1/2}\xrightarrow{} \rm5^2{P}_{3/2}$). To lock the repetition rate $\rm\Omega_{Rep}$ of the SMC, the seed laser is modulated using an intensity modulator (IM) by a radio frequency (RF) reference $\rm\Omega_{RF}$. Then the beatnote between the first-order side bands of the SMC and the RF-modulated seed laser is detected using a photodetector (PD) with 1-GHz bandwidth. Benefiting from the PD bandwidth limitation, only the difference frequency $\rm\Delta\Omega_r=\Omega_{RF}-\Omega_{Rep}$ is passively filter out and used as the error signal to lock $\rm\Omega_{Rep}$ through feeding back the operation temperature of the MRR. The repetition rate is stabilized to the frequency of the local RF reference when $\rm\Delta\Omega_r$ approaches to zero. 

Figure \ref{fig2:soliton}b shows the optical spectra of two independent SMC chips, which could provide more than 200 channels of frequency teeth with SNR higher than 20 dB. The stability of repetition rates ($\rm\Delta\Omega_r$) of both SMC chips are shown in Fig. \ref{fig2:soliton}c and \ref{fig2:soliton}d. The peak-to-peak and standard deviation of repetition rate fluctuations are only 1442 Hz and 215 Hz, respectively, within 3 hours. Figures \ref{fig2:soliton}e-\ref{fig2:soliton}f present the beat frequencies $\rm\Delta\Omega_{teeth}$ and their statistical distributions of three pairs of comb teeth (CH0\&CH0, CH+15\&CH+15, and CH-15\&CH-15, corresponding to C21\&C21, C14\&C14 and H28\&H28 of the standard ITU channels, respectively) within the concurrent time period. The corresponding standard deviations of beat frequencies are 30.7 kHz, 31.1 kHz and 29.6 kHz, respectively. The standard deviations of beat frequencies are limited by the characteristic linewidth of the ${}^{87}$\textbf{Rb} atomic transition process \cite{leopold2016_rb_locking}. The enlarged view in Fig. \ref{fig2:soliton}e shows no statistically significant difference across these tooth pairs. The remarkable homogeneity of beat frequencies observed provides compelling evidence for the consistency in repetition rates between these independent SMC chips. 

The peak-to-peak thermal tuning range is merely less than 150 mK in 3 hours (Figs. \ref{fig2:soliton}g-\ref{fig2:soliton}h) for the active repetition-rate stabilization. It is significantly smaller than the SMC survival temperature range of 2K \cite{huang2025_soliton_hom}. Compared to microwave injection locking, active temperature feedback can effectively reduce the intracavity power, mitigating the influence of background waves generated by injected sidebands on soliton formation. Additionally, active locking offers stronger resistance to environmental perturbations \cite{niur2024_soliton_locking}. 

The locally frequency-locking to the same but independent references significantly simplifies the network complexity, making the large-scale network more robust for practical deployment. The long-term frequency-locking precision and stability of comb teeth, enabled by locally fully locked SMCs, satisfy the stringent requirements of a high-performance and large-scale massively parallel MDI-QKD network. \cite{weikj2020_chip_mdiqkd}.

\subsection{Transmitter chip characterizations and HOM interferences}
\label{subsection:QKDchip}
\begin{figure}[htbp!] 
	\centering
	\includegraphics[width=\textwidth]{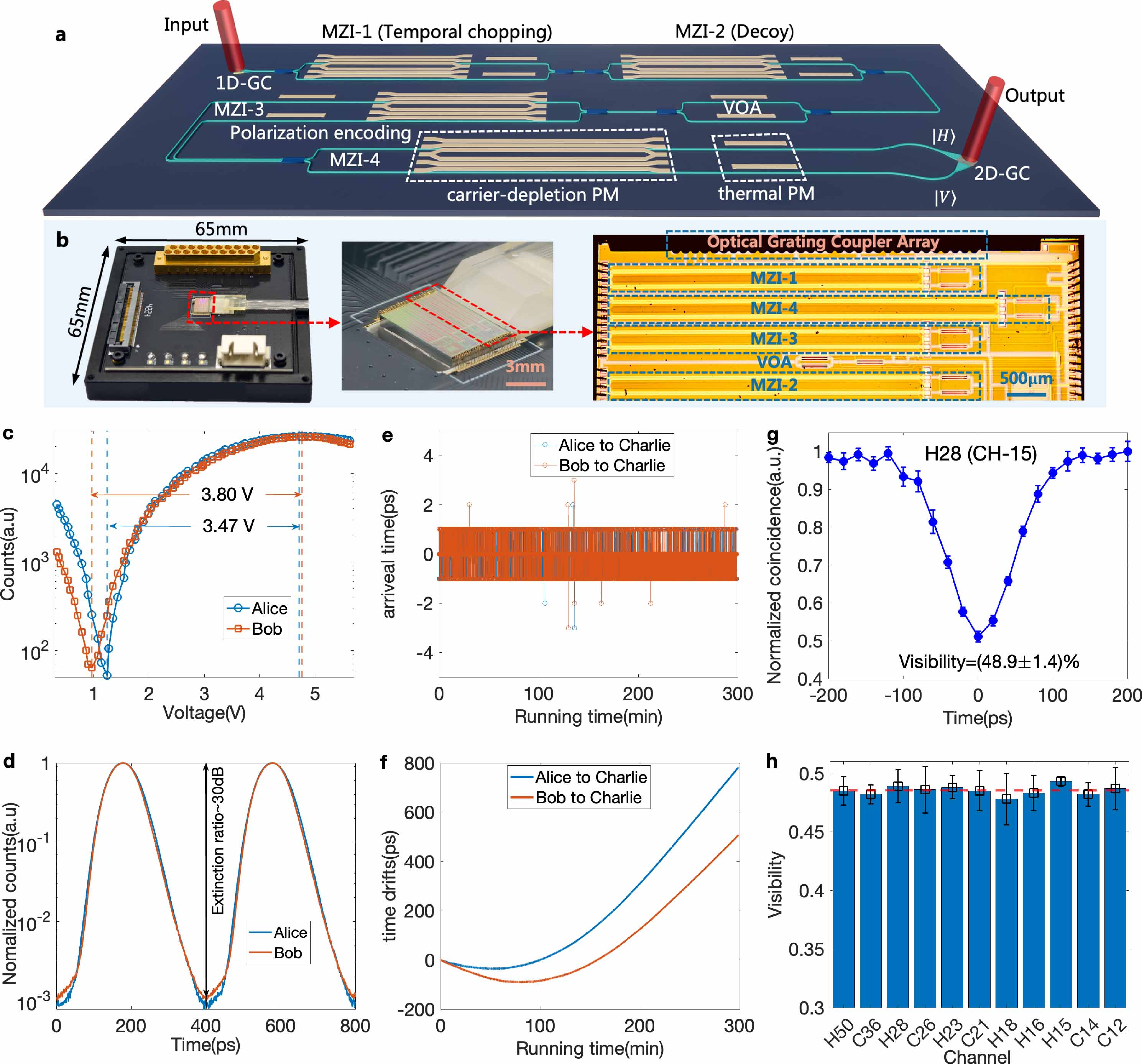} 	
	\caption{\textbf{Transmitter chip characterization.}
		\textbf{a}, Schematic of the polarization-encoded transmitter chip (4 MZIs + VOA). \textbf{b}, The images of the optoelectronic packaged transmitter chip. \textbf{c}, Voltage-dependent transmittance of the MZIs of both transmitter chips. \textbf{d}, The normalized temporal intensity profiles of the pulsed comb teeth modulated by the on-chip MZI, which includes the time jitter of the SNSPD. Arrival time drifts of photon pulses \textbf{e}, with  and \textbf{f}, without time-drift compensation after transmission over a 100-km fiber channel. \textbf{g}, HOM interference fringe of the 90-ps tooth pair H28\&H28. \textbf{h}, HOM visibilities of tooth pairs from H50\&H50 to C12\&C12 (corresponding to CH-60\&CH-60 to CH+19\&CH+19 of the comb teeth).}
	\label{fig3:chip_performance} %
\end{figure}

High-performance integrated transmitter chips are indispensable to realize low-cost and large-scale fully connected quantum network. The silicon-photonic-chip-based MDI-QKD transmitters are operated at 2.5 GHz. As shown in Fig. \ref{fig3:chip_performance}a, the transmitter chip incorporates four electro-optic and one thermal-optic Mach-Zehnder inteferometers (MZIs) in a monolithic design. Both of carrier-depletion and thermal tuning phase modulators (PMs) are employed for the electro-optic MZIs. The optical carrier is coupled into and out of the transmitter chip through 1D and 2D grating couplers (1D-GC and 2D-GC), respectively. MZI-1 to MZI-4 operate in push-pull modulation mode, where MZI-1/2 chop the CW optical carrier into pulsed signals and realize decoy-state intensity modulation. The subsequent thermal-optic MZI serves as a variable optical attenuator (VOA) to achieve single-photon level output. MZI3 and MZI-4 jointly prepare the mutually unbiased polarization bases $Z$ and $X$. The upper (down) path of MZI-4 is mapped to $|H\rangle (|V\rangle)$ via the 2D-GC. High-speed switch between $|H\rangle$ and $|V\rangle)$ is achieved through carrier-depletion PMs in MZI-4. Figure \ref{fig3:chip_performance}b illustrates the compact optoelectronic packaging ($\rm65\times65\times16{mm}^3$) of the transmitter chip, with magnified views showing the photonic integration details.

Figure \ref{fig3:chip_performance}c displays the voltage-dependent modulation characteristics of the MZIs at 2.5-GHz operation rate, where the half-wave voltages are 3.47 V for Alice and 3.80 V for Bob. As presented in Fig. \ref{fig3:chip_performance}d, the optical carriers are chopped into 95-ps pulses (FWHM) through MZI-1, with 30-dB extinction ratios. The propagation of narrow photon pulses through the long-distance fiber channel is subject to dynamic temporal shifts, which impairs two-photon interference visibility. The photon arrival time is monitored and compensated in real time. 
It effectively suppresses arrival time drifts of Alice$\rightarrow$Charlie and Bob$\rightarrow$Charlie to $\sim$2 ps (Fig. \ref{fig3:chip_performance}e), which is two orders of magnitude smaller compared to that without compensation (Fig. \ref{fig3:chip_performance}f).

Based on the high-performance SMC and transmitter chips, high-visibility HOM interferences between Alice and Bob are realized using the 95-ps photon pulses, where a multi-channel superconductor single-photon detector (SNSPD) is used and the count rate per channel is about 1 Mcps. Figure \ref{fig3:chip_performance}g gives the HOM interference dip of H28\&H28. The HOM interference visibility is $(48.9\pm1.4)\%$, approaching the theoretical limitation of 50.0\%. Figure \ref{fig3:chip_performance}h displays HOM interference visibilities for teeth pairs spanning H50\&H50 to C12\&C12. All tooth pairs covering 80 channels exhibit consistent interference quality and the average visibility is as high as 48.5\% (red dashed line). See \textbf{supplementary materials} for detailed HOM interferences of these channels.

\subsection{Secure key sharing rate}
\label{subsection:MDI-QKD}
\begin{figure}[htbp!] 
	\centering
	\includegraphics[width=\textwidth]{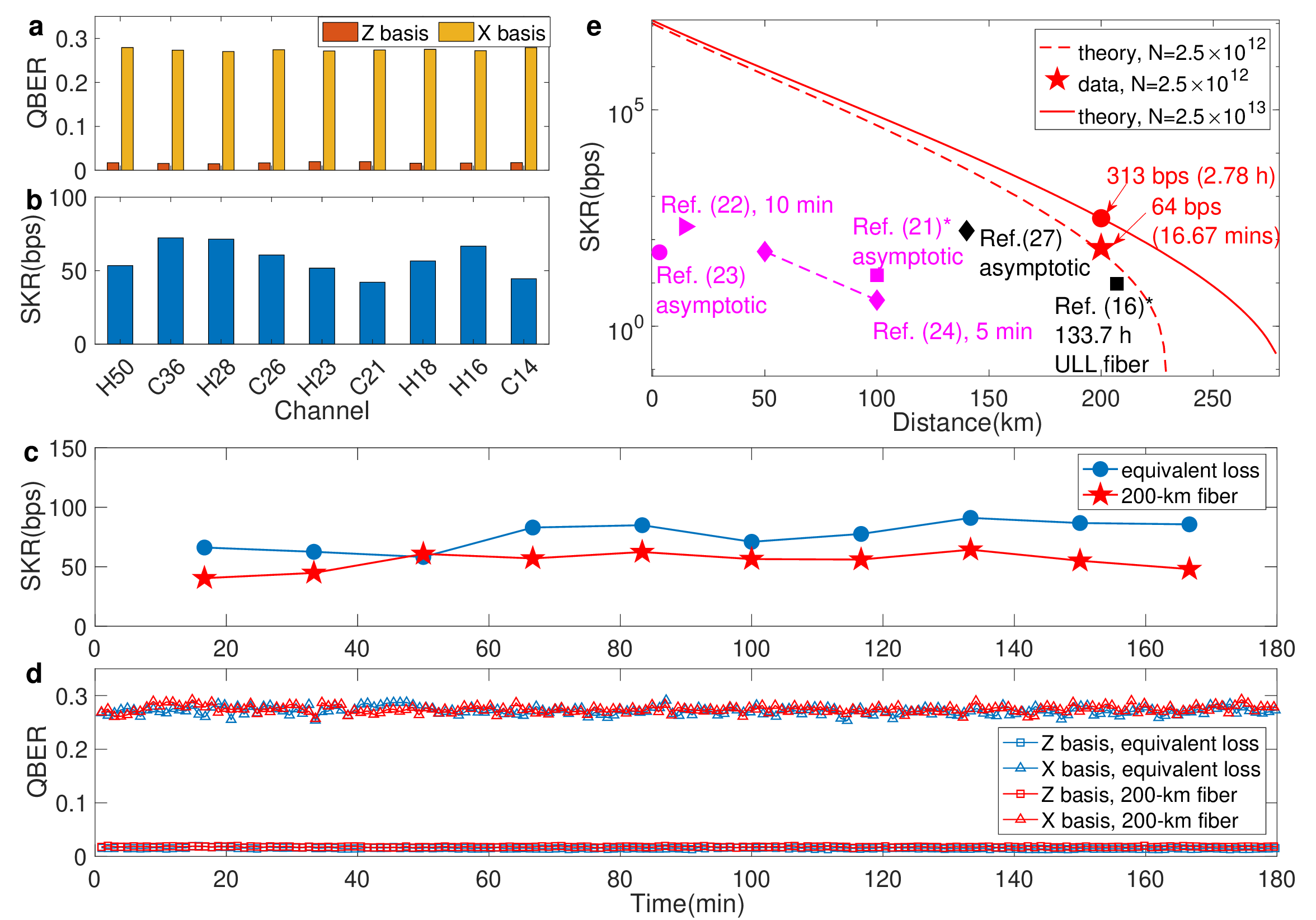} 	
	\caption{\textbf{Secure key rates and QBERs of integrated parallel MDI-QKD system over 200 km fiber.}
		\textbf{a}, QBERs and \textbf{b}, secure key rates of different tooth pairs spanning H50\&H50 to C14\&C14. \textbf{c-d}, Real-time secure key rates and QBERs of H28\&H28 over 3 hours' running, where secure keys is extracted every 1000 s. The blue and red colors denote the equivalent loss and 200-km fiber channels, respectively. \textbf{e}, Secure key rates per channel as distance. ${}^*$: Ref. \cite{yinhl2016_404km_mdi} used ultralow loss (ULL, 0.16 dB/km) fiber and Ref. \cite{wengerowsky2018_entanglement_fully_network} simulated the channel with attenuators.}
	\label{fig4:SKR}
\end{figure}

By utilizing the highly stable SMC and high SNR encoding transmitter chips, we achieve massively parallel MDI-QKD over 200-km ﬁber channel with high secure key rate, where no trusted quantum network provider is required. The polarization-based MDI-QKD adopts four-intensity decoy state method. The high frequency-uniformity of SMC and 30-dB-ER transmitter chips ensure the low QBERs of quantum states preparation. Combining with the stochastic parallel gradient descent (SPGD) algorithm \cite{vorontsov1998_spgd}, electrically driven polarization controllers (EPCs) are used to dynamically compensate the polarization drifts over the 200-km fiber channels (see \textbf{supplementary materials} for details).

Figures \ref{fig4:SKR}a and \ref{fig4:SKR}b show the QBERs and secure key rates of different MDI-QKD channels using comb teeth pairs spanning H50\&H50 to C14\&C14 (covering 76 channels). The secure key rate is calculated by considering finite-size effect \cite{jiangc_2021_mdi_skr}, where the failure probability is limited to $<1\times10^{-10}$ and the accumulation time for each key extraction round is 1000 s ($\sim$16.67 minutes). The average QBER and secure key rate of $Z$ ($X$) basis of all channels are $(1.7\pm0.2)\%$ ($(27.4\pm0.3)\%$) and 58 bps, separately with a finite sending pulses number of 1000 s ($\sim$16.67 minutes). See \textbf{supplementary materials} for detailed discussion of decoy state parameters and finite-size effect analysis.

Figures \ref{fig4:SKR}c and \ref{fig4:SKR}d show the longtime running secure key rates and QBERs. The system works as well as that within 1000 s and demonstrates the good-performance ability of the dynamical polarization compensation. The secure key rate and QBERs are smooth for both equivalent-loss and 200-km standard single-mode fiber channels. And the system attains a secure key rate as high as 64 bps over 200-km fiber channel. The consistence of performances with equivalent-loss and 200-km-fiber channels demonstrates the good stability of our system for long-distance and large-scale fully connected quantum network with untrusted network providers.

Figure \ref{fig4:SKR}e compares the secure key rates of the proposed fully connected quantum network with previous advances. Remarkably, the key rate per channel here is much higher, while the data accumulation time interval to extract a finite-size key block of our system is much shorter than the 10$\sim$133 hours of previous single-channel discrete-device \cite{yinhl2016_404km_mdi,woodward2021_mdi_1ghz} and chip-based MDI-QKD systems \cite{weikj2020_chip_mdiqkd}. By increasing the accumulation time to 10000 s ($\sim$2.78 hours), the secure key rate per channel will become 313 bps at 200 km (red solid line), which is 2-order of magnitude higher than previous record with 200-km ultralow loss (ULL) fiber \cite{yinhl2016_404km_mdi}. On the other hand, when comparing to the previous fully connected quantum networks without rigorous finite-size effect analysis \cite{wengerowsky2018_entanglement_fully_network,joshi2020_entanglement_fully_network,liux2022_entanglement_fully_network,huangy2025_entanglement_fully_network}, the key sharing rate and distance of the proposed architecture are 3-4 orders of magnitude higher and 100 km further, respectively. 

\subsection{Discussion and outlook}
\label{subsection:discussion}

We demonstrate a pioneering fully connected quantum network architecture that simultaneously achieves high-speed and large-scale deployment performance through the high-precision SMC manipulation and low-QBER quantum states preparation. The integration of both optical SMCs and QKD transmitters within compact optoelectronic package chips enables substantial reduction in infrastructure expenditure, while maintaining network-provider-independent quantum-grade security across all network links, representing a significant milestone in the practical implementation of quantum communication network. 

The operation channels from H50 to C14 are experimental verified for massively parallel fully connected MDI-QKD network. Actually, the number of channels is solely limited by the passband range of the AWGs used in our experiments. Considering the performance-homogeneity of the frequency locked SMC chips and the ultra-low crosstalks between these massive comb teeth (see \textbf{supplementary materials} for details), the proposed architecture can support a fully connect quantum network with thousands of users, combining with both frequency ($\sim$200 channels) and time division multiplexing techniques. It is 2 orders of magnitude larger than previous implementations.

Our work is the first long-distance polarization-encoded MDI-QKD that achieves finite-size secure keys, as the previous work only evaluated asymptotic secure key rate over 140-km fiber \cite{weikj2020_chip_mdiqkd}. By developing a dynamic polarization compensation strategy, we achieve a mean secure key rate of 64 bps per channel (extracted every 16.67 minutes) over 200-km fiber for long-time running. It significantly benefits real-time root key update, while the key update time intervals of previous MDI-QKD were 10-133 hours \cite{yinhl2016_404km_mdi,weikj2020_chip_mdiqkd}. It removes the obstacles for establishing an intercity fully connected MDI-QKD network. As the locking uncertainty of pump laser and repetition rate of the SMC are only 31 kHz and 215 Hz, respectively, our SMC source meets the frequency stability requirement of mode-pairing protocols and can further increase the secure key rate significantly \cite{xieym2022_mpqkd,zengp2022_mpqkd}. 

Remarkably, the proposed network architecture can also be modified simply to a quantum-repeater-based quantum network by taking the coherent comb teeth as pump lasers to generate entangled photon pairs and realizing entanglement swapping with untrusted quantum relays. Therefore, the architectural paves the way for the deployment of high-speed, large-scale fully connected quantum networks involving untrusted network providers across metropolitan and intercity regions. It not only enhances the practicality of deployment but also provides a novel design perspective for the secure fully connected quantum network.


\clearpage 

\bibliography{MDIQKD.bib} 

@article{Bennett2014_BB84,
	author = {Bennett, Charles H. and Brassard, Gilles},
	journal = {Theoretical Computer Science},
	month = {dec},
	pages = {7--11},
	pmid = {20355177},
	publisher = {IEEE},
	title = {{Quantum cryptography: Public key distribution and coin tossing}},
	volume = {560},
	year = {2014}
}

@INPROCEEDINGS{GLLP2004,
	author={Gottesman, D. and Lo, H.-K. and Lutkenhaus, N. and Preskill, J.},
	booktitle={International Symposium onInformation Theory, 2004. ISIT 2004. Proceedings.}, 
	title={Security of quantum key distribution with imperfect devices}, 
	year={2004},
	volume={},
	number={},
	pages={136},
}

@article{WangXB2005_decoy,
	author = {Wang, Xiang-Bin},
	journal = {Physical Review Letters},
	month = {jun},
	number = {23},
	pages = {230503},
	title = {{Beating the Photon-Number-Splitting Attack in Practical Quantum Cryptography}},
	volume = {94},
	year = {2005}
}

@article{Lo2005_decoy,
	author = {Lo, Hoi-Kwong and Ma, Xiongfeng and Chen, Kai},
	journal = {Physical Review Letters},
	month = {jun},
	number = {23},
	pages = {230504},
	title = {{Decoy State Quantum Key Distribution}},
	volume = {94},
	year = {2005}
}

@article{XuFH2020_RMP,
	author = {Xu, Feihu and Ma, Xiongfeng and Zhang, Qiang and Lo, Hoi-Kwong and Pan, Jian-Wei},
	journal = {Reviews of Modern Physics},
	month = {may},
	number = {2},
	pages = {025002},
	publisher = {American Physical Society},
	title = {{Secure quantum key distribution with realistic devices}},
	volume = {92},
	year = {2020}
}

@article{Lucamarini2018_TF,
	author = {Lucamarini, M. and Yuan, Z. L. and Dynes, J. F. and Shields, A. J.},
	journal = {Nature},
	month = {may},
	number = {7705},
	pages = {400--403},
	title = {{Overcoming the rate–distance limit of quantum key distribution without quantum repeaters}},
	volume = {557},
	year = {2018}
}

@article{Minder2019_TF,
	author = {Minder, M. and Pittaluga, M. and Roberts, G. L. and Lucamarini, M. and Dynes, J. F. and Yuan, Z. L. and Shields, A. J.},
	journal = {Nature Photonics},
	month = {may},
	number = {5},
	pages = {334--338},
	publisher = {Springer US},
	title = {{Experimental quantum key distribution beyond the repeaterless secret key capacity}},
	volume = {13},
	year = {2019}
}

@article{WangS2022_TF,
	author = {Wang, Shuang and Yin, Zhen-Qiang and He, De-Yong and Chen, Wei and Wang, Rui-Qiang and Ye, Peng and Zhou, Yao and Fan-Yuan, Guan-Jie and Wang, Fang-Xiang and Chen, Wei and Zhu, Yong-Gang and Morozov, Pavel V and Divochiy, Alexander V and Zhou, Zheng and Guo, Guang-Can and Han, Zheng-Fu},
	journal = {Nature Photonics},
	month = {feb},
	number = {2},
	pages = {154--161},
	title = {{Twin-field quantum key distribution over 830-km fibre}},
	volume = {16},
	year = {2022}
}

@article{LiuY2023_1000km,
	author = {Liu, Yang and Zhang, Wei-Jun and Jiang, Cong and Chen, Jiu-Peng and Ma, Di and Zhang, Chi and Pan, Wen-Xin and Dong, Hao and Xiong, Jia-Min and Zhang, Cheng-Jun and Li, Hao and Wang, Rui-Chun and Lu, Chao-Yang and Wu, Jun and Chen, Teng-Yun and You, Lixing and Wang, Xiang-Bin and Zhang, Qiang and Pan, Jian-Wei},
	journal = {Quantum Frontiers},
	month = {nov},
	number = {1},
	pages = {16},
	title = {1002 km twin-field quantum key distribution with finite-key analysis},
	volume = {2},
	year = {2023}
}

@article{Frohlich2013_BB84_network,
author = {Fr{\"{o}}hlich, Bernd and Dynes, James F. and Lucamarini, Marco and Sharpe, Andrew W. and Yuan, Zhiliang and Shields, Andrew J.},
journal = {Nature},
month = {sep},
number = {7465},
pages = {69--72},
title = {{A quantum access network}},
volume = {501},
year = {2013}
}

@article{WangS2014_BB84_network,
author = {Wang, Shuang and Chen, Wei and Yin, Zhen-Qiang and Li, Hong-Wei and He, De-Yong and Li, Yu-Hu and Zhou, Zheng and Song, Xiao-Tian and Li, Fang-Yi and Wang, Dong and Chen, Hua and Han, Yun-Guang and Huang, Jing-Zheng and Guo, Jun-Fu and Hao, Peng-Lei and Li, Mo and Zhang, Chun-Mei and Liu, Dong and Liang, Wen-Ye and Miao, Chun-Hua and Wu, Ping and Guo, Guang-Can and Han, Zheng-Fu},
journal = {Optics Express},
month = {sep},
number = {18},
pages = {21739--21756},
title = {{Field and long-term demonstration of a wide area quantum key distribution network}},
volume = {22},
year = {2014}
}

@article{ChenYA2021_space-to-ground_network,
author = {Chen, Yu Ao and Zhang, Qiang and Chen, Teng Yun and Cai, Wen Qi and Liao, Sheng Kai and Zhang, Jun and Chen, Kai and Yin, Juan and Ren, Ji Gang and Chen, Zhu and Han, Sheng Long and Yu, Qing and Liang, Ken and Zhou, Fei and Yuan, Xiao and Zhao, Mei Sheng and Wang, Tian Yin and Jiang, Xiao and Zhang, Liang and Liu, Wei Yue and Li, Yang and Shen, Qi and Cao, Yuan and Lu, Chao Yang and Shu, Rong and Wang, Jian Yu and Li, Li and Liu, Nai Le and Xu, Feihu and Wang, Xiang Bin and Peng, Cheng Zhi and Pan, Jian Wei},
journal = {Nature},
month = {jan},
number = {7841},
pages = {214--219},
title = {{An integrated space-to-ground quantum communication network over 4,600 kilometres}},
volume = {589},
year = {2021}
}

@article{Hong1987_HOM,
author = {Hong, C. K. and Ou, Z. Y. and Mandel, L.},
journal = {Physical Review Letters},
month = {nov},
number = {18},
pages = {2044--2046},
pmid = {4162727},
title = {{Measurement of subpicosecond time intervals between two photons by interference}},
volume = {59},
year = {1987}
}

@article{Lo2012_MDI,
author = {Lo, Hoi-Kwong and Curty, Marcos and Qi, Bing},
journal = {Physical Review Letters},
month = {mar},
number = {13},
pages = {130503},
title = {{Measurement-Device-Independent Quantum Key Distribution}},
volume = {108},
year = {2012}
}

@article{Braunstein2012_MDI,
author = {Braunstein, Samuel L and Pirandola, Stefano},
journal = {Physical Review Letters},
month = {mar},
number = {13},
pages = {130502},
title = {{Side-channel-free quantum key distribution}},
volume = {108},
year = {2012}
}

@article{TangYL2016_MDI_network,
author = {Tang, Yan-Lin and Yin, Hua-Lei and Zhao, Qi and Liu, Hui and Sun, Xiang-Xiang and Huang, Ming-Qi and Zhang, Wei-Jun and Chen, Si-Jing and Zhang, Lu and You, Li-Xing and Wang, Zhen and Liu, Yang and Lu, Chao-Yang and Jiang, Xiao and Ma, Xiongfeng and Zhang, Qiang and Chen, Teng-Yun and Pan, Jian-Wei},
journal = {Physical Review X},
month = {mar},
number = {1},
pages = {011024},
title = {{Measurement-Device-Independent Quantum Key Distribution over Untrustful Metropolitan Network}},
volume = {6},
year = {2016}
}

@article{Roberts2017_MDI_network,
author = {Roberts, G. L. and Lucamarini, M. and Yuan, Z. L. and Dynes, J. F. and Comandar, L. C. and Sharpe, A. W. and Shields, A. J. and Curty, M. and Puthoor, I. V. and Andersson, E.},
journal = {Nature Communications},
month = {oct},
number = {1},
pages = {1098},
title = {{Experimental measurement-device-independent quantum digital signatures}},
volume = {8},
year = {2017}
}

@article{WangC2022_MDI_network,
author = {Wang, Chao and Kon, Wen Yu and Ng, Hong Jie and Lim, Charles C.-W.},
journal = {Light: Science \& Applications},
month = {sep},
number = {1},
pages = {268},
publisher = {Optica Publishing Group},
title = {{Experimental symmetric private information retrieval with measurement-device-independent quantum network}},
volume = {11},
year = {2022}
}

@article{YinHL2016_404km_MDI,
author = {Yin, Hua-Lei and Chen, Teng-Yun and Yu, Zong-Wen and Liu, Hui and You, Li-Xing and Zhou, Yi-Heng and Chen, Si-Jing and Mao, Yingqiu and Huang, Ming-Qi and Zhang, Wei-Jun and Chen, Hao and Li, Ming Jun and Nolan, Daniel and Zhou, Fei and Jiang, Xiao and Wang, Zhen and Zhang, Qiang and Wang, Xiang-Bin and Pan, Jian-Wei},
journal = {Physical Review Letters},
month = {nov},
number = {19},
pages = {190501},
title = {{Measurement-Device-Independent Quantum Key Distribution Over a 404 km Optical Fiber}},
volume = {117},
year = {2016}
}

@article{Woodward2021_MDI_1GHz,
author = {Woodward, R. I. and Lo, Y. S. and Pittaluga, M. and Minder, M. and Para{\"{i}}so, T. K. and Lucamarini, M. and Yuan, Z. L. and Shields, A. J.},
journal = {npj Quantum Information},
month = {apr},
number = {1},
pages = {58},
title = {{Gigahertz measurement-device-independent quantum key distribution using directly modulated lasers}},
volume = {7},
year = {2021}
}

@article{Fan-Yuan2021_MDI_network,
author = {Fan-Yuan, Guan-Jie and Lu, Feng-Yu and Wang, Shuang and Yin, Zhen-Qiang and He, De-Yong and Zhou, Zheng and Teng, Jun and Chen, Wei and Guo, Guang-Can and Han, Zheng-Fu},
journal = {Photonics Research},
month = {oct},
number = {10},
pages = {1881},
title = {{Measurement-device-independent quantum key distribution for nonstandalone networks}},
volume = {9},
year = {2021}
}

@article{Pittaluga2025_TF_network,
author = {Pittaluga, Mirko and Lo, Yuen San and Brzosko, Adam and Woodward, Robert I and Scalcon, Davide and Winnel, Matthew S and Roger, Thomas and Dynes, James F and Owen, Kim A and Ju{\'{a}}rez, Sergio and Rydlichowski, Piotr and Vicinanza, Domenico and Roberts, Guy and Shields, Andrew J},
journal = {Nature},
month = {apr},
number = {8060},
pages = {911--917},
title = {{Long-distance coherent quantum communications in deployed telecom networks}},
volume = {640},
year = {2025}
}

@article{Wengerowsky2018_entanglement_fully_network,
author = {Wengerowsky, S{\"{o}}ren and Joshi, Siddarth Koduru and Steinlechner, Fabian and H{\"{u}}bel, Hannes and Ursin, Rupert},
journal = {Nature},
number = {7735},
pages = {225--228},
publisher = {Springer US},
title = {{An entanglement-based wavelength-multiplexed quantum communication network}},
volume = {564},
year = {2018}
}

@article{Joshi2020_entanglement_fully_network,
author = {Joshi, Siddarth Koduru and Aktas, Djeylan and Wengerowsky, S{\"{o}}ren and Lon{\v{c}}ari{\'{c}}, Martin and Neumann, Sebastian Philipp and Liu, Bo and Scheidl, Thomas and Lorenzo, Guillermo Curr{\'{a}}s and Samec, {\v{Z}}eljko and Kling, Laurent and Qiu, Alex and Razavi, Mohsen and Stip{\v{c}}evi{\'{c}}, Mario and Rarity, John G and Ursin, Rupert},
journal = {Science Advances},
month = {sep},
number = {36},
pages = {eaba0959},
title = {{A trusted node–free eight-user metropolitan quantum communication network}},
volume = {6},
year = {2020}
}

@article{LiuX2022_entanglement_fully_network,
author = {Liu, Xu and Liu, Jingyuan and Xue, Rong and Wang, Heqing and Li, Hao and Feng, Xue and Liu, Fang and Cui, Kaiyu and Wang, Zhen and You, Lixing and Huang, Yidong and Zhang, Wei},
journal = {PhotoniX},
month = {dec},
number = {1},
pages = {2},
title = {40-user fully connected entanglement-based quantum key distribution network without trusted node},
volume = {3},
year = {2022}
}

@article{HuangY2025_entanglement_fully_network,
author = {Huang, Yiwen and Qi, Zhantong and Yang, Yilin and Zhang, Yuting and Li, Yuanhua and Zheng, Yuanlin and Chen, Xianfeng},
journal = {Laser \& Photonics Reviews},
month = {jan},
number = {1},
pages = {2301026},
title = {{A Sixteen‐user Time‐bin Entangled Quantum Communication Network With Fully Connected Topology}},
volume = {19},
year = {2025}
}

@article{Sibson2017_chip_BB84QKD,
author = {Sibson, P. and Erven, C. and Godfrey, M. and Miki, S. and Yamashita, T. and Fujiwara, M. and Sasaki, M. and Terai, H. and Tanner, M. G. and Natarajan, C. M. and Hadfield, R. H. and O'Brien, J. L. and Thompson, M. G.},
journal = {Nature Communications},
month = {feb},
number = {1},
pages = {13984},
title = {{Chip-based quantum key distribution}},
volume = {8},
year = {2017}
}

@article{WangJW2018_chip_quantum_computing,
author = {Wang, Jianwei and Paesani, Stefano and Ding, Yunhong and Santagati, Raffaele and Skrzypczyk, Paul and Salavrakos, Alexia and Tura, Jordi and Augusiak, Remigiusz and Man{\v{c}}inska, Laura and Bacco, Davide and Bonneau, Damien and Silverstone, Joshua W and Gong, Qihuang and Ac{\'{i}}n, Antonio and Rottwitt, Karsten and Oxenl{\o}we, Leif K and O'Brien, Jeremy L. and Laing, Anthony and Thompson, Mark G},
journal = {Science},
number = {6386},
pages = {285--291},
title = {{Multidimensional quantum entanglement with large-scale integrated optics}},
volume = {360},
year = {2018}
}

@article{WeiKJ2020_chip_MDIQKD,
author = {Wei, Kejin and Li, Wei and Tan, Hao and Li, Yang and Min, Hao and Zhang, Wei-Jun and Li, Hao and You, Lixing and Wang, Zhen and Jiang, Xiao and Chen, Teng-Yun and Liao, Sheng-Kai and Peng, Cheng-Zhi and Xu, Feihu and Pan, Jian-Wei},
journal = {Physical Review X},
month = {aug},
number = {3},
pages = {031030},
publisher = {American Physical Society},
title = {{High-Speed Measurement-Device-Independent Quantum Key Distribution with Integrated Silicon Photonics}},
volume = {10},
year = {2020}
}

@article{Paraiso2021_chip_BB84QKD,
author = {Para{\"{i}}so, Taofiq K. and Roger, Thomas and Marangon, Davide G. and {De Marco}, Innocenzo and Sanzaro, Mirko and Woodward, Robert I. and Dynes, James F. and Yuan, Zhiliang and Shields, Andrew J.},
journal = {Nature Photonics},
number = {11},
pages = {850--856},
publisher = {Springer US},
title = {{A photonic integrated quantum secure communication system}},
volume = {15},
year = {2021}
}

@article{LiW2023_chip_BB84QKD,
author = {Li, Wei and Zhang, Likang and Tan, Hao and Lu, Yichen and Liao, Sheng-Kai and Huang, Jia and Li, Hao and Wang, Zhen and Mao, Hao-Kun and Yan, Bingze and Li, Qiong and Liu, Yang and Zhang, Qiang and Peng, Cheng-Zhi and You, Lixing and Xu, Feihu and Pan, Jian-Wei},
journal = {Nature Photonics},
month = {may},
number = {5},
pages = {416--421},
title = {{High-rate quantum key distribution exceeding 110 Mb s$^{–1}$}},
volume = {17},
year = {2023}
}

@article{Herr2014_soliton,
author = {Herr, T. and Brasch, V. and Jost, J. D. and Wang, C. Y. and Kondratiev, N. M. and Gorodetsky, M. L. and Kippenberg, T. J.},
journal = {Nature Photonics},
month = {feb},
number = {2},
pages = {145--152},
publisher = {Nature Publishing Group},
title = {{Temporal solitons in optical microresonators}},
volume = {8},
year = {2014}
}

@article{Kippenberg2018_soliton_review,
author = {Kippenberg, Tobias J. and Gaeta, Alexander L. and Lipson, Michal and Gorodetsky, Michael L.},
journal = {Science},
month = {aug},
number = {6402},
pages = {eaan8083},
title = {{Dissipative Kerr solitons in optical microresonators}},
volume = {361},
year = {2018}
}

@article{stern2018_onchip_soliton,
author = {Stern, Brian and Ji, Xingchen and Okawachi, Yoshitomo and Gaeta, Alexander L. and Lipson, Michal},
journal = {Nature},
number = {7727},
pages = {401--405},
publisher = {Springer US},
title = {{Battery-operated integrated frequency comb generator}},
volume = {562},
year = {2018}
}

@article{Marin-Palomo2017_optical_communication,
author = {Marin-Palomo, Pablo and Kemal, Juned N. and Karpov, Maxim and Kordts, Arne and Pfeifle, Joerg and Pfeiffer, Martin H. P. and Trocha, Philipp and Wolf, Stefan and Brasch, Victor and Anderson, Miles H. and Rosenberger, Ralf and Vijayan, Kovendhan and Freude, Wolfgang and Kippenberg, Tobias J. and Koos, Christian},
journal = {Nature},
month = {jun},
number = {7657},
pages = {274--279},
publisher = {Nature Publishing Group},
title = {{Microresonator-based solitons for massively parallel coherent optical communications}},
volume = {546},
year = {2017}
}

@article{ShuHW2022_soliton_optical_communication,
author = {Shu, Haowen and Chang, Lin and Tao, Yuansheng and Shen, Bitao and Xie, Weiqiang and Jin, Ming and Netherton, Andrew and Tao, Zihan and Zhang, Xuguang and Chen, Ruixuan and Bai, Bowen and Qin, Jun and Yu, Shaohua and Wang, Xingjun and Bowers, John E.},
journal = {Nature},
month = {may},
number = {7910},
pages = {457--463},
title = {{Microcomb-driven silicon photonic systems}},
volume = {605},
year = {2022}
}

@article{Wangfx2020_soliton_QKD,
author = {Wang, Fang-Xiang and Wang, Weiqiang and Niu, Rui and Wang, Xinyu and Zou, Chang‐Ling and Dong, Chun‐Hua and Little, Brent E. and Chu, Sai T. and Liu, Hang and Hao, Penglei and Liu, Shufeng and Wang, Shuang and Yin, Zhen‐Qiang and He, De‐Yong and Zhang, Wenfu and Zhao, Wei and Han, Zheng‐Fu and Guo, Guang‐Can and Chen, Wei},
journal = {Laser \& Photonics Reviews},
month = {feb},
number = {2},
pages = {1900190},
title = {{Quantum Key Distribution with On‐Chip Dissipative Kerr Soliton}},
volume = {14},
year = {2020}
}

@article{Huang2025_soliton_HOM,
author = {Huang, Long and Wang, Weiqiang and Wang, Fangxiang and Wang, Yang and Zou, Changling and Tang, Linhan and Little, Brent E and Zhao, Wei and Han, Zhengfu and Yang, Jun and Wang, Guochao and Chen, Wei and Zhang, Wenfu},
journal = {Science Advances},
month = {jan},
number = {5},
pages = {eadq8982},
title = {{Massively parallel Hong-Ou-Mandel interference based on independent soliton microcombs}},
volume = {11},
year = {2025}
}

@article{NiuR2024_soliton_locking,
author = {Niu, Rui and Wan, Shuai and Hua, Tian-Peng and Wang, Wei-Qiang and Wang, Zheng-Yu and Li, Jin and Wang, Zhu-Bo and Li, Ming and Shen, Zhen and Sun, Yu Robert and Hu, Shui-Ming and Little, Brent E. and Chu, Sai Tak and Zhao, Wei and Guo, Guang-Can and Zou, Chang-Ling and Xiao, Yun-Feng and Zhang, Wen-Fu and Dong, Chun-Hua},
journal = {Science China Physics, Mechanics \& Astronomy},
month = {feb},
number = {2},
pages = {224262},
title = {{Atom-referenced and stabilized soliton microcomb}},
volume = {67},
year = {2024}
}

@article{Leopold2016_Rb_locking,
author = {Leopold, T. and Schm{\"{o}}ger, L. and Feuchtenbeiner, S. and Grebing, C. and Micke, P. and Scharnhorst, N. and Leroux, I. D. and L{\'{o}}pez-Urrutia, J. R. Crespo and Schmidt, P. O.},
journal = {Applied Physics B},
month = {sep},
number = {9},
pages = {236},
publisher = {Springer Berlin Heidelberg},
title = {{A tunable low-drift laser stabilized to an atomic reference}},
volume = {122},
year = {2016}
}

@article{Jiangc_2021_MDI_SKR,
author = {Jiang, Cong and Yu, Zong-Wen and Hu, Xiao-Long and Wang, Xiang-Bin},
journal = {Physical Review A},
month = {jan},
number = {1},
pages = {012402},
title = {{Higher key rate of measurement-device-independent quantum key distribution through joint data processing}},
volume = {103},
year = {2021}
}

@article{XieYM2022_MPQKD,
author = {Xie, Yuan-Mei and Lu, Yu-Shuo and Weng, Chen-Xun and Cao, Xiao-Yu and Jia, Zhao-Ying and Bao, Yu and Wang, Yang and Fu, Yao and Yin, Hua-Lei and Chen, Zeng-Bing},
journal = {PRX Quantum},
month = {apr},
number = {2},
pages = {020315},
title = {{Breaking the Rate-Loss Bound of Quantum Key Distribution with Asynchronous Two-Photon Interference}},
volume = {3},
year = {2022}
}

@article{ZengP2022_MPQKD,
author = {Zeng, Pei and Zhou, Hongyi and Wu, Weijie and Ma, Xiongfeng},
journal = {Nature Communications},
month = {jul},
number = {1},
pages = {3903},
title = {{Mode-pairing quantum key distribution}},
volume = {13},
year = {2022}
}

@article{Vorontsov1998_SPGD,
author = {Vorontsov, M. A. and Sivokon, V. P.},
journal = {Journal of the Optical Society of America A},
month = {oct},
number = {10},
pages = {2745},
title = {{Stochastic parallel-gradient-descent technique for high-resolution wave-front phase-distortion correction}},
volume = {15},
year = {1998}
}
\bibliographystyle{sciencemag}


\section*{Acknowledgments}
\paragraph*{Funding:}
This work is supported by National Natural Science Foundation of China (grant nos. 62371437 and 62075238), Quantum Science and Technology-National Science and Technology Major Project (Grant No. 2021ZD0300701) and CAS Project for Young Scientists in Basic Research (grant no. YSBR-069).

\paragraph*{Author contributions:}
W.C. and F.X.W. proposed the original idea; F.X.W., W.C., W.Z., W.W. and C.L.Z. developed the detailed proposal; F.X.W., S.T.Z, G.W.Z. carried out the QKD and network experiments and analyzed the data with the assistance of G.S.W., W.J.D., Z.H.W and S.W.; G.W.Z., F.X.W. and W.C. designed and carried out the QKD transmitter chips; B.E.L., L.H, W.W. and W.Z. designed and carried out the SMC chips; L.H., G.S.W., W.W. and W.Z. developed the SMC locking technique with the assistance of G.W. and L.Z.; Z.H.W. and Z.Q.Y. developed the theoretical security model of MDI-QKD; F.X.W., S.T.Z., G.W.Z. and W.C. wrote the manuscript, and W.W., L.H. and W.Z. revised the manuscript; All authors discussed and approved the data; W.C., W.Z., G.C.G. and Z.F.H. supervised the project.

\paragraph*{Competing interests:}
The authors declare that they have no competing interests.
\paragraph*{Data and materials availability:}
All data needed to evaluate the conclusions in the paper are present in the paper and/or the Supplementary Materials.

\end{document}